\documentstyle[aps,prl,twocolumn,epsfig]{revtex}
\newcommand{\be}{\begin{equation}} 
\newcommand{\ee}{\end{equation}} 
\newcommand{\erf}{\mbox{erf}}
\begin{document}
\twocolumn[\hsize\textwidth\columnwidth\hsize\csname@twocolumnfalse\endcsname
\title {Low autocorrelated multi-phase sequences}
\author{ Liat Ein-Dor$^1$,Ido Kanter$^1$,  
and Wolfgang Kinzel$^2$}
\address{
     $^1$ Department of Physics, Bar-Ilan University\\ 
       \mbox{~} Ramat-Gan, 52900 Israel \\       
     $^2$ Institut f\"{u}r Theoretische Physik, Universit\"{a}t W\"{u}rzburg\\
      \mbox{~} Am Hubland, D-97074 W\"{u}rzburg, Germany
}
\maketitle

\begin{abstract}
The interplay between the ground state energy of the generalized 
Bernasconi model to multi-phase,
and the minimal value of the maximal autocorrelation function, 
$C_{max}=\max_K{|C_K|}$, $K=1,..N-1$, is examined  
analytically and the main results are: 
%liat
(a) The minimal value of $\min_N{C_{max}}$ is $0.435\sqrt{N}$ 
    significantly smaller than the typical value for random sequences 
    $O(\sqrt{\log{N}}\sqrt{N})$.
(b) $\min_N{C_{max}}$ over all sequences of length $N$ is 
obtained in an energy which  
is about $30\%$ above the ground-state energy of the generalized
Bernasconi model, 
independent of the number of phases $m$.
 (c) The maximal merit factor $F_{max}$ grows linearly with $m$.
 (d) For a given $N$, $\min_N{C_{max}}\sim\sqrt{N/m}$ 
    indicating that for $m=N$,
    $\min_N{C_{max}}=1$, i.e. a Barker code exits.   
The analytical results are confirmed by simulations. 

\end{abstract}
]
In many applications of communication science \cite{Schroder,Shapiro}, 
as well as a variety of other fields, it is necessary to find 
sequences with low autocorrelation functions.
Some of these applications utilize the pulse compression feature
of the low autocorrelated sequences to obtain high resolution in 
radars and sonars. In other applications, the shifts of 
such periodic sequences can be used to identify users in multi-user
 systems.  
Due to their importance, low autocorrelated sequences 
have evoked a wide spread interest accompanied
 by the development of 
various methods for constructing such sequences \cite{Schroder}.

In order to construct a binary sequence $S=(s_1,...,s_N)$ with low off
 peak autocorrelations, one has to define which quantity has to 
be minimized.
The different applications divide the low autocorrelated sequences
into two types in correspondence with the quantity which they minimize. 
The first kind of low autocorrelated sequences minimizes  
 the Hamiltonian of the Bernasconi model
 \cite{Bernasconi,Krauth,Mezard,Parisi} which is given by:

\be
H=\sum_{K=1}^{N-1}C_{K}^2
\label {energy}
\ee
where for the case of non-periodic boundary conditions, which is at 
the center of our study, 
\be
C_{K}=\sum_{j=1}^{N-K}s_js_{j+K},
\label {CK}
\ee
and for the periodic case
$C_{K}=\sum_{j=1}^{N}s_js_{((j+K-1)\bmod N)+1}$.
Note that for random sequences the average value of $H$ 
in the non-periodic case is $N^2/2$,
whereas for low autocorrelated sequences this energy is reduced
by a merit factor $F>1$ to $N^2/2F$. 
However, there are applications for which   
the maximal off peak autocorrelation, $C_{max}=\max_K{|C_K|}$,
 has to be minimized. The second kind of low autocorrelated sequences 
are the solutions of this minimization problem. 
%liat 
Note that for random sequences $C_{max}$ is typically  
$O(\sqrt{\log{N}}\sqrt{N})$. 
For a sequence of length $N$, 
the maximal possible ratio between the peak, $|C_0|=N$,
and the maximal off-peak autocorrelations, 
$|C_K|$ with $K=1,...,N-1$ is $N/1$.
The only known binary sequences with this ratio are the Barker codes of length
$2$, $3$, $4$, $5$, $7$, $11$ and $13$. Obviously, the Barker sequences, 
when they exist, furnish a minimum for the two minimization problems.
However, Turyn \cite{Turyn} has shown that no other 
binary codes such as this exist for any length less than $144$ or for odd 
length greater than $13$.
An exact solution for these two minimization problems 
is known only for systems (up to $N=59$)
 which are small enough 
to permit an effective exhaustive search \cite{Mertens1,Mertens2}.
 Extrapolation of the ground state energies which were 
found for small systems using an exhaustive search 
indicates
$F_{max}=\lim_{N\to\infty}\frac{N^2}{2H}=8.5$ \cite{Mertens1}. Moreover,
 $F_{max}$ was conjectured by Golay 
\cite{golay1,golay2} to be bounded
 from above by $12.324$.

The following questions regarding low autocorrelated sequences
 are still open and are at the center of our study: (a) Do 
the sequences that minimize ${C_{max}}$ minimize 
    the energy of the Bernasconi model as well?  
(b) In case that the two minimization problems are not equivalent, how
    far are the energy values of the sequences which minimize ${C_{max}}$
    from the values of the ground state energy?   
(c) Exhaustive search of sequences with length up to $N=59$ show that 
    the degeneracy of the ground state is of O(1) and is bounded from 
    above by $8$. Is the degeneracy of the states which minimize 
    $C_{max}$ is of O(1) as well? 
(d) In this context, which of the two minimal quantities is easier
    to approach?
Additional interesting questions arise when multi-phase 
sequences, whose terms are complex $m$-th roots of $1$
for $m>2$, are considered. For such sequences 
\be
s_l=\exp(i\frac{2\pi}{m}l) 
\label {s_l}
\ee
where l=1,...,m and the correlations are defined as 
\be
C_{K}=\sum_{j=1}^{N-K}s_j{s^*_{j+K}}.
\ee
where $s^*_{j+K}$ is the conjugate of $s_{j+K}$. 
The information about multi-phase low autocorrelated sequences 
is more limited, mainly since the configuration space grows 
like $m^N$ which makes the exhaustive search ineffective even 
for very small sequences. Hence, not much is known
about the influence of 
the number of phases both on the 
maximal value of $F$ and on the minimal value of $C_{max}$.
Is it possible to attain better solutions by increasing the 
number of phases?

\begin{figure}
%\centerline{\epsfxsize=3.25in   \epsffile{Bernfig1.eps}}
\narrowtext \epsfxsize=3in
\hskip 0.5in
\epsfig{width=0.9\columnwidth,file=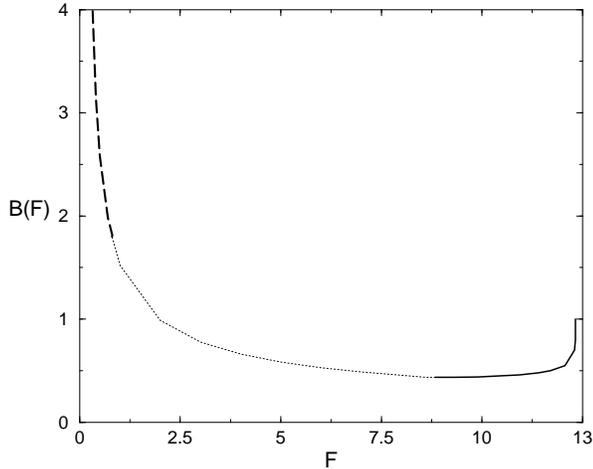}
\vskip 0.15in
\caption{$B(F)$ vs. $F$. This graph is composed of three regions
corresponding to the three different behaviours of $P_F(C_{max})$.
In the dashed region $0.215\leq F\leq 0.8$, 
the fraction of sequences $P_F(C_{max})$ 
is getting larger as $F$ 
increases, while in the dotted-line region, $0.8<F<F_c$,  
this fraction decreases 
to $2^{-N}$ as $F$ approaches $F_c$. 
For $F>F_c$ (solid-line) 
 the fraction of sequences 
with the minimal value 
$B(F)\sqrt{N}$ is constant and equals $2^{-N}$.}
\label {fig1}
\end{figure}

\begin{figure}
\narrowtext \epsfxsize=3in 
\hskip 0.5in
\epsfig{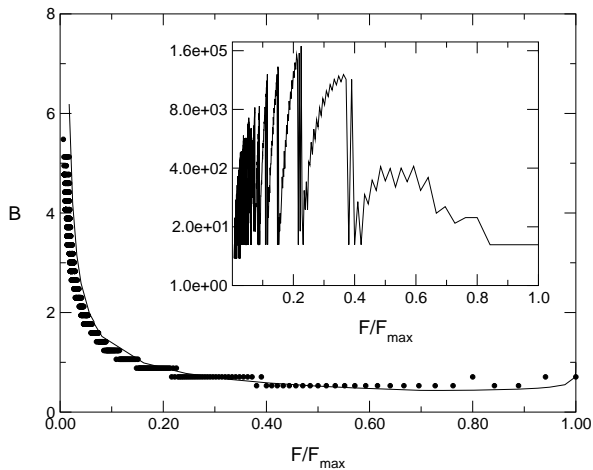}
\vskip 0.1in
\caption{$B$ vs. $F/F_{max}$. The solid line represents 
the analytical results while the circles represent the exhaustive search 
results for $N=32$. Note that the analytical $F_{max}$ is $12.324$ whereas 
for $N=32$ $F_{max}=8$.
Inset: a histogram describing the logarithm of the number of sequences with 
the minimal upper bound $B$ for each of the micro-canonical ensembles
 with a merit factor $F$ as was derived from an exhaustive search
for $N=32$.}
\label {fig2}
\end{figure}

In this study we suggest an analytical technique to calculate the 
quantities of interest, namely, $\max_{N}F$ and $\min_{N}{C_{max}}$
for both binary and multi-phase sequences.
An obvious lower bound of $\min_{N,F}{C_{max}}$ over
 all sequences of length $N$ and a merit factor 
$F$ is obtained by assuming that all $|C_K|$ equal $\sqrt{N/(2F)}$.
Nevertheless it is tempting to investigate the relations between $F$
and $\min_{N,F}{C_{max}}$ and to measure more accurately 
the optimal value $\min_{N}{C_{max}}$ over all sequences of length $N$
similarly to the aforementioned upper bound given for $F_{max}$ \cite{golay1}.
 For our analytical study we consider the non-periodic 
Bernasconi model. We approximate the sequences to be random
 and the autocorrelation functions to be independent variables
 with the following Gaussian distribution
\be
P(C_{K})=\frac{1}{\sqrt{2\pi(N-K)}}\exp(\frac{-C_{K}^2}{2(N-K)}).
\label {P(C)}
\ee
Using Eq. (\ref {P(C)}), the probability $P_F(C_{max})$ that the 
autocorrelations of a sequence with a merit factor 
$F=N^2/2H$ have an upper bound $C_{max}$ such that $|C_K|\le {C_{max}}$
is given by
\be
P_F(C_{max})=\int_{-C_{max}}^{C_{max}}\prod_K{dC_K}
P(C_{K})\delta(\frac{1}{N}\sum_{K=1}^{N-1}{C_K^2-
\frac{N}{2F}}) .
\label {P_F}
\ee
Since there are $2^N$ distinct binary sequences of length $N$, 
it is necessary that $P_F\ge{2^{-N}}$ in order for a sequence with the 
corresponding features, namely a merit factor $F$ 
and an upper bound $C_{max}$,
to exist. Hence, by equating $P_F(C_{max})$ to $2^{-N}$ one can 
find the minimal upper bound $\min_{N,F}{C_{max}}$ among the upper bounds
of all sequences with a merit factor $F$.
Moreover, the Gaussian distributions of $C_K$ imply that 
 $\min_{N,F}{C_{max}}$ must be of $O(\sqrt{N})$ in order for 
$P_F(C_{max})$ to be equal 
$2^{-N}$. Assigning 
$\min_{N,F}{C_{max}}=B(F)\sqrt{N}$ and inserting the integral representation 
of 
$\delta$ function in 
Eq. (\ref {P_F}) , the saddle point method can be used to obtain the
 following set of equations
\begin{eqnarray}
\frac{\lambda}{2F}-\frac{1}{4\lambda}((1-2\lambda)\ln(1-2\lambda)+2\lambda)
\nonumber \\
-\int_0^{1}\ln{\erf(B(F)g(y,\lambda))}dy=\ln{2},
\label {A_min1}
\end{eqnarray}

\begin{eqnarray}
\frac{1}{2F}-\frac{1}{4\lambda^2}(\ln(1-2\lambda)+2\lambda)+
\nonumber \\
B(F)\sqrt{\frac{2}{\pi}}\int_0^{1}{{\exp(-B(F)^2
g(y,\lambda)^2)}\over{
\erf(B(F)g(y,\lambda))\sqrt{2}g(y,\lambda)}}dy=0.
\label {A_min2}
\end{eqnarray}
\noindent
where $g(y,\lambda)=\sqrt{\frac{1-2\lambda(1-y)}{2(1-y)}}$. 
Solving numerically this set of equations, it turns out that a solution 
exists only for a bounded region of $F$, where $.215\leq F\leq 12.324$. 
These two limit values of $F$ are alternatively obtained by 
taking the limits of the integration in Eq. (\ref {P_F})    
to infinity.
The existence of a lower bound is a consequence of the fact that as the energy 
becomes smaller the allowed values of the correlations decrease 
and the probability for such sequences (Eq. (\ref {P_F}))
 decreases as well. As was pointed out above, when this probability drops 
below $2^{-N}$, there are no more sequences with the required energy.
Similarly, for high energy values, high values of correlations are 
required and the corresponding probabilities become smaller again.
The analysis of $B(F)$ inside the allowed region reveals 
that the minimal value of $B(F_c)=0.435$ is obtained for $F_c=8.839$ 
whereas
the maximal merit factor $F=12.324$ 
corresponds to a higher value of $B(F)=1$. Thus, the 
merit factor of sequences which minimizes $C_{max}$ is around $71\%$
lower then $F_{max}$. 
Exhaustive search results in the region $N\in [15,28]$
indicate that the average ratio between 
the two merit factors, $<<<F(\min{C_{max}})>>/F_{max}>_N$, 
is $\sim 0.6771$ where $<<$ $>>$ denotes the average
over all sequences which minimize ${C_{max}}$ whereas $<$ $>_N$ denotes
the average over the different sequences length $N$.
The deviation from the analytical result $\sim 0.717$
is attributed  
to finite size effects.
Fig. \ref{fig1} shows the behavior of $B(F)$ derived from the numerical 
solution of Eqs. (\ref {A_min1}) and (\ref {A_min2}).
The behavior of $P_F(C_{max})$ divides this graph into three regimes.
Starting with the large $F$ regime, there is only
 one sequence whose $C_{max}$ equals $B(F)\sqrt{N}$.
This feature holds until $F_c=8.839$ for which
the minimal $B(F)=0.435$ is obtained. 
Further decrease of $F$ results in an increasing number 
of sequences
with an upper bound of $B(F)\sqrt{N}$. This increment terminates at
 $F=0.8$ and for smaller values of $F$ the number of sequences 
reduces to $1$ as $F$ approaches $0.215$.  
In order to compare the analytical behaviour of $B$
to that obtained from simulations, it is necessary to find a way
to circumvent the mismatch between the analytical values of $F$ and 
those obtained from simulations.
The rescaling of $F$ by $F_{max}$ yields a parameter $F/F_{max}$
whose range of values, $[0,1]$, does not depend on a specific realization.
This allows a comparison between analytical results and simulations.
Results of an exhaustive search over sequences of length $N=32$
show that the behavior of $B$ as a function of $F/F_{max}$ resembles
the analytical prediction (Fig. \ref{fig2}). 

Moreover, the existence 
of the three aforementioned regions
of $P_F(C_{max})$ is confirmed by counting the number of sequences with 
the minimal upper bound $B$ for each of the micro-canonical ensembles
 with a merit factor $F$ and length $N=32$.
The histogram is demonstrated in the inset of Fig. \ref{fig2}, where
the merit factor is rescaled to $F/F_{max}$.
In the small $F$ regime, the average number of sequences increases
with $F$ until $F/F_{max}\sim 0.2$ from which the number 
of sequences decreases up to the plateau which starts at 
$F/F_{max}\sim 0.83$.

%\vspace {0.1cm}

\begin{figure}
%\centerline{\epsfxsize=3.25in   \epsffile{Bernfig3.eps}}
%\vspace{0.4cm}
\narrowtext \epsfxsize=3in
\hskip 0.5in
\epsfig{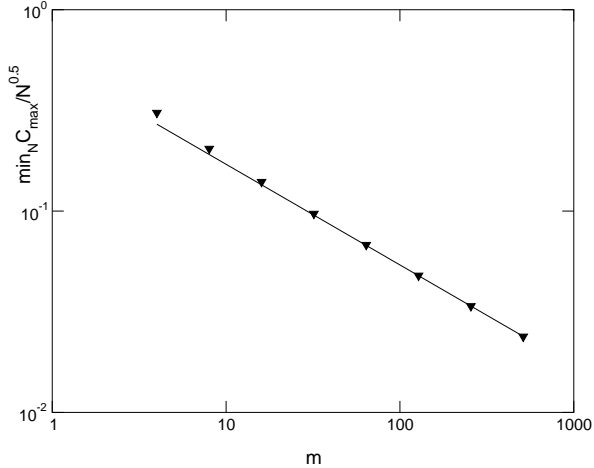}
\vskip 0.15in
\caption{ 
$\min_N{C_{max}}/\sqrt{N}$ vs. $m$ as was 
obtained from the numerical solution of Eqs. 
(\ref {B_min1}) and (\ref {B_min2}).
 The solid line is the least square fit $0.54/\sqrt{m}$.}
\label {fig3}
\end{figure}

\begin{figure}
%\epsfxsize= 0.49\textwidth
%\epsffile{Berninset45.eps}
\narrowtext \epsfxsize=3in
\hskip 0.5in
\epsfig{width=0.9\columnwidth,file=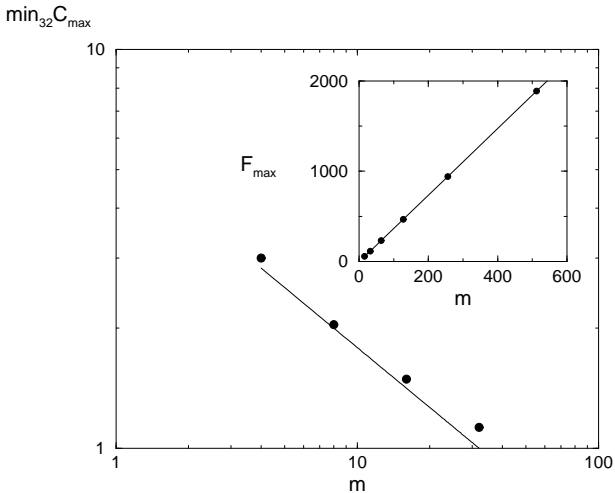}
\vskip 0.15in
\caption{Results for $\min_{N}{C_{max}}$ 
as a function of $m$ for sequences of length $N=32$.
The filled circles stand for the simulated annealing results.
The deviation from the analytical results (solid-line) is attributed to
finite size effects and to the suboptimal solution 
obtained in our limited 
running times of the simulations.
Inset: Analytical results for $F_{max}$ as a function of $m$ 
which were obtained from the numerical solution of Eqs. (\ref {B_min1})
 and (\ref {B_min2}). The solid line is the least square fit, $3.7m$.}
\label {fig45}
\end{figure}

Now we turn to study the influence of the number of phases both on the 
maximal value of $F$ and on the minimal value of $C_{max}$.
We assign $D_K$ and $E_K$ to be independent Gaussian variables which 
represent the real and the imaginary parts of $C_K$
respectively. In case that $m=2^p$ where $p$ is an integer number greater
than $1$,
 these two parts have the same probability 
distribution
\be
P(D_{K})=\frac{1}{\sqrt{\pi(N-K)}}\exp(\frac{-D_{K}^2}{(N-K)}).
\label {P_m(D)}
\ee
Under the same assumptions that have been used in the derivation 
of the binary case,
Eq. (\ref {P_F}) becomes 
\begin{eqnarray}
P^m_F(C_{max})=\int_{-C_{max}}^{C_{max}}\prod_K{dD_KdE_K}
P(D_{K})P(E_{K})
\nonumber \\
\delta(\frac{1}{N}\sum_{K=1}^N{(D_K^2+E_K^2)-\frac{N}{2F}}).
\label {P_F(B)}
\end{eqnarray}
Similarly to the binary case,
 if $P^m_F(C_{max})=m^{-N}$ 
then $B(F)\sqrt{N}$ 
is the minimal value of $C_{max}$ for which
there is a sequence with a merit factor $F$.
The same procedure that has been used to calculate $B(F)$
for the binary case results in 
the following set of equations
\begin{eqnarray}
\frac{\lambda}{2F}-\frac{1}{\lambda}((1-\lambda)\ln(1-\lambda)+\lambda)
\nonumber \\
-2\int_0^{1}\ln{\erf(B(F)g(y,\lambda))}dy&=&\ln (m).
\label {B_min1}
\end{eqnarray}

\begin{eqnarray}
\frac{1}{2F}-\frac{1}{\lambda^2}(\ln(1-\lambda)+\lambda)+   
\nonumber \\   
\frac{2B(F)}{\sqrt{\pi}}\int_0^{1}{{
\exp(-B(F)^2g(y,\lambda)^2)}\over{
\erf(B(F)g(y,\lambda))g(y,\lambda)}}dy=0,
\label {B_min2}
\end{eqnarray}
where $g(y,\lambda)=\sqrt{\frac{1-\lambda(1-y)}{(1-y)}}$.
Solving numerically these two equations,
the behaviors of $\min_N{C_{max}}/\sqrt{N}$ and $F_{max}$ as a function of $m$
are obtained. Fig.  \ref{fig3} shows that $\min_N{C_{max}}/\sqrt{N}$
is proportional to $1/\sqrt{m}$ with slight deviations for small $m$.
This relation between $B(F_c)=\min_N{C_{max}}/\sqrt{N}$ and $m$ implies
that such a relation holds for $B(F_{max})$ as well.
Since the variance of $C_K^2$ equals $N-K$, 
their typical values drop linearly with $K$. 
Hence, $H=N^2/2F$ is approximated by a sum of an algebric series of $N-1$  
terms, $C_K^2$, 
with an upper bound, $C_{max}^2$, of $O(N/m)$. 
Assuming that $C_K^2$ is homogeneously distributed
 between $0$ to $C_{max}^2$ yields a linear increment of 
$F_{max}$ as a function of $m$ in agreement with the numerical solution 
of $F_{max}$ which is depicted in the inset of Fig. \ref{fig45}.   
The results of $F_{max}$ and $B(F_c)=\min_N{C_{max}}/\sqrt{N}$ 
as a function of $m$ show that
the solutions of the two minimization problems are improved by
increasing the number of phases $m$. 

\vspace{4mm}

\vspace*{-3mm}

These results raise the 
question whether it is possible to increase $m$ such that 
 $\min_N{C_{max}}$ 
becomes 1. An asymptotic expansion of Eqs. (\ref {B_min1})
and (\ref {B_min2}) reveals that for $m=N$, $\min_N{C_{max}}=1$.
Note that $C_{N-1}$ is always 1 and therefore $\min_N{C_{max}}=1$ 
holds for the entire regime $m>N$.

%$C_K$ is a sum of $N-K$ terms of magnitude $1$ with $m$
% possible directions in the complex space. Since $m=2^p$,
% each of the $m$ possible vectors has an inverse vector 
%in its opposite direction. Hence when $N-K$ is odd,
% $C_K$ is bound from below by $1$ corresponding to optimal cases
% in which pairs of terms in 
%opposite directions are cancelled and one single term is left. 
%Consequently, the result $\min_N{C_{max}}=1$ holds for the entire 
%regime $m>N$.
We used the simulated annealing
method for sequences of lengths $N=32$ 
with different number of phases $m$, to find $\min_N{C_{max}}$.
The relatively small sequences size was chosen to 
enable an appropriate scan of the configuration space
 in reasonable computational time. 
The results are exhibited in Fig. \ref{fig45}, 
and support the anticipated $\sqrt{N/m}$ behavior of $\min_N{C_{max}}$.
For $m=32$, there is a deviation of $\min_N{C_{max}}$
from the analytical prediction
 $\min_N{C_{max}}=1$, probably 
since the simulated annealing method yields only 
suboptimal solutions.

Finally, we would like to examine the simulated annealing method in 
light of the results of this study. Simulations show that 
for the same running times, the simulated 
annealing method yields ${C_{max}}$ which is closer to its 
minimal value than $\sum_{K=1}^{N-1}C_K^2$.  
This can be explained by the larger degeneracy of ${C_{max}}$
compared with that of $\sum_{K=1}^{N-1}C_K^2$.
Moreover, it turns out that in order to minimize ${C_{max}}$,
it is preferable to start the searching process with the minimization 
of the energy function, $\sum_{K=1}^{N-1}C_K^2$,  
and then replace it with ${C_{max}}$ \cite {unpublished}. In this way 
the system avoids the plateaus which characterize the landscape 
of $C_{max}$ in the configuration space.
However, future research is necessary to find out how the 
results of this study can be applied to further improve the
 searching processes
of low autocorrelated sequences.

Critical comments on the manuscript by R. Metzler are acknowledged.
The research is supported in part by the Israel Academy of science.

\end{document}